\documentclass[12pt]{article}
\input epsf

\newcommand{\VEV}[1]{\left\langle{#1}\right\rangle}

\newcommand{\ket}[1]{\,\left|\,{#1}\right\rangle}
\renewcommand{\bar}[1]{\overline{#1}}
\newcommand{\etal} {{\em et al.}}
\newcommand{\cf}{{\em c.f.}}
\newcommand{\ie}  {{\em i.e.}}
\newcommand{\eg} {{\em e.g.}}
\newcommand{\M} {{\cal M}}
\newcommand{\R} {{\cal R}}
\newcommand{\inn}{\nonumber}

\newcommand{\eqcm}{\: ,}
\newcommand{\eqpt}{\: .}
\newcommand{\as}{\alpha_s}
\newcommand{\lqcd}{\Lambda_{QCD}}

\newcommand{\dpt}{\Delta_\perp}

\newcommand{\pom}{{\cal P}}
\newcommand{\odd}{{\cal O}}
\newcommand{\eq}[1]{Eq.~(\ref{#1})}

\begin{document}

\begin{flushright}
{\small
SLAC--PUB--8198\\
July 1999\\}
\end{flushright}

\vfill
\begin{center}
{\bf\large   
PERSPECTIVES ON EPIC PHYSICS\footnote{\baselineskip=13pt
Work supported by the Department
of Energy, contract DE--AC03--76SF00515.}}

\medskip

Stanley J. Brodsky        \\
{\em Stanford Linear Accelerator Center \\
 Stanford Univerity, Stanford, California 94309\\
email: sjbth@slac.stanford.edu} \\
\medskip

\end{center}

\vfill

\begin{center}
{\bf\large   
Abstract }
\end{center}

An electron-proton/ion polarized
beam collider (EPIC) with high luminosity and center of mass energy $\sqrt
s = 25$
GeV would be a valuable facility for
fundamental studies of proton
and nuclear structure and tests of quantum chromodynamics,
I review a sample
of prospective EPIC topics,  particularly
semi-exclusive reactions,
studies of the proton fragmentation
region, heavy quark electroproduction, and a new probe of odderon/pomeron
interference.

\vfill

\begin{center}
Talk presented at the \\
EPIC'99 Workshop \\
Indiana University Cyclotron Facility \\
Bloomington, Indiana \\
April 8--11, 1999
\end{center}
\vfill
\newpage

\section{Introduction}

A central goal of both high energy and nuclear physics is to unravel the
structure and dynamics
of nucleons and nuclei in terms of their fundamental quark and gluon
degrees of freedom.  An
outstanding option for such experimental studies is an
electron-proton/ion polarized beam collider (EPIC) in an intermediate
energy domain
well above the fixed
target facilities available at Jefferson Laboratory or SLAC, and well below
the high energy range of
the electron-proton collider HERA at DESY.  For
example, an electron beam of
$4$ GeV colliding with protons or ions at $40$ GeV would provide
center-of-mass energies $\sqrt s
\cong 25$ GeV, well above both the open charm and bottom thresholds and
well-matched to QCD studies.  The equivalent electron
laboratory beam energy is $E^{e^-}_{\rm Lab} \cong 300$ GeV.  The EPIC
collider is envisioned to
have polarized beams and high luminosity, ${\cal L} \cong 10^{33}{\rm
cm}^{-2} {\rm sec}^{-1}$.
With good duty factor for studying exclusive final states and
mulitiparticle correlations, and with full angular
acceptance,---especially in the beam fragmentation region---the EPIC
facility would constitute a complete ``electron
microscope" for testing QCD and illuminating hadron and
nuclear substructure.

Our present empirical knowledge of the quark and gluon distributions of the
proton has revealed
a remarkably complex substructure.  It is helpful to categorize parton
distributions as
``intrinsic" --pertaining to the composition of the target hadron, and
``extrinsic", reflecting
the resolved PQCD substructure of the individual quarks and gluons
themselves. For example, If sea quarks were generated
solely by gluon splitting, the anti-quark distributions would be isospin
symmetric. However, the
$\bar u(x)$ and
$\bar d(x)$ antiquark distributions of the proton at $Q^2
\sim 10$ GeV$^2$ have quite different shapes, which must reflect dynamics
intrinsic to the proton's structure.  We now know that gluons carry a
significant fraction of the proton's spin as well as its momentum.  Since
gluon exchange between
valence quarks contributes to the
$p-\Delta$ mass splitting, it follows that the gluon distributions
cannot be solely accounted for by  bremsstrahlung from
individual quarks.
Similarly,  in the case of
heavy quarks, $s\bar s$,
$c \bar c$, $b \bar b$, the diagrams in which the sea quarks are
multiply-connected to
the proton's valence quarks are intrinsic to the proton's structure
itself.  Thus neither gluons nor sea quarks are solely
generated by DGLAP evolution, and one cannot define a resolution scale
in momentum transfer in deep inelastic scattering $Q_0$ where the sea or
gluon degrees of freedom can be neglected.  There are also remarkable
surprises associated with
the chirality distributions of the quarks
$\Delta q = q_{\uparrow/\uparrow} - q_{\downarrow/\uparrow}$ of the valence
quark
$\Delta u(x,Q)$ and $\Delta d(x,Q)$, which again show that a simple
valence quark
approximation to nucleon spin structure functions is
far from the actual dynamical
situation.

An electron-proton/ion collider in the EPIC energy regime would
clearly be a valuable facility for many types of QCD
studies.  A large array of such topics
were discussed in this workshop. Examples include:

\begin{enumerate}
\item
The polarized
beams of EPIC would
provide the capability for studying detailed spin and azimuthal
correlations, reflecting the helicity distributions of the quark and gluon
constituents and the physics of spin transfer
to the final state
hadrons in the jet beam fragmentation region.\cite{Mulders:1999yt}

\item
The study of the proton fragmentation regime;
\ie\ the observation of the remnants of the disassociated proton left behind
after  a
struck is struck. Veneziano and Trentadue \cite{Trentadue:1994ka} have
emphasized the proton's ``fracture functions" arising from the evolution of the
target fragments with
$Q^2$. More generally,  the physics of the fragmentation region, even
Coulomb dissociation,  reflects the actual
composition of the light-cone wave functions of hadrons and nuclei,  a largely
unexplored area of QCD which could provide important insights into the
fundamental structure of the nucleon.

\item
The study of the interface of coherent and incoherent quark phenomena.
This includes the
large $x\rightarrow 1$ regime where perturbative QCD predicts specific
power-law fall off of scattering amplitudes
and one expects duality between exclusive and inclusive channels.  Further,
one can study
``semi-exclusive'' reactions such as $\gamma^* p \rightarrow \pi^+ X$ where
the meson
is produced at large transverse momentum in isolation of other hadrons, \ie\
a large rapidity
gap.\cite{acw,Brodsky:1998sr}  Such reactions provide the capability of
designing
effective currents which
probe specific parton distributions.

\item
The study of purely exclusive reactions, such as
large momentum transfer fixed-angle reactions,
meson photoproduction, processes
which all depend in detail on  hadron distribution amplitudes
$\phi_H(x_i,Q)$, the basic wavefunctions of the hadrons.\cite{BL,Efremov:1980qk}

\item
The study of diffractive processes where the interacting proton or nucleus
remains intact,
such as deeply virtual Compton scattering,
$\gamma^* p \rightarrow \gamma p$, diffractive meson electroproduction
$\gamma^* p \rightarrow \phi p$,
$\gamma^* p \rightarrow \pi^0 p$ processes which
are sensitive to the dynamics of the QCD hard pomeron and odderon, the
$C=+1$, and $-1$
exchange systems which control high energy scattering.  Measurements of
diffractive
dijet production $\gamma^* p \rightarrow$ Jet Jet $p$ test the QCD
structure of virtual
photons.  The charge asymmetry in $\gamma^* p \rightarrow c\bar c p$
measures
pomeron/odderon interference and has high sensitivity
to the odderon amplitude.\cite{Brodsky:1999mz}

\item
The nuclear beam capability of the EPIC collider is important for
identifying basic and novel
features of fundamental nuclear dynamics, such as color transparency,
hidden color, and
specific dynamics associated with the non-additive ``EMC'' features of
nuclear structure
functions.  Color transparency\cite{BM,Frankfurt:1991rk} reflects the fact that
hadrons are fluctuating systems, and
that only the small transverse size valence quark components of the hadron
wavefunctions enter virtual photoproduction processes such as $\gamma^* A
\rightarrow
\rho^0 A^\prime$, and hard quasi-elastic processes such as $eA \rightarrow
e^\prime p (A-1)$.
Such compact wavefunctions are predicted to have minimal shadowing and other
hadronic interactions.   The first evidence
for QCD ``color transparency" was observed in quasi-elastic $p p$
scattering in nuclei.\cite{heppelmann90} In contrast to color
transparency, Fock states with large-scale color configurations
interact strongly and with high particle number production.
\cite{bbfhs93}

\item
Hidden color is a fundamental prediction of QCD,
reflecting the
fact that the nuclear wavefunction cannot be solely described as composite
of nucleon
color singlet configurations.\cite{bjl83}  Such components should show up in
non-additivity of nuclear
amplitudes as well as the complex structure of the nuclear fragmentation region.

\item
The energy range of the EPIC facility would allow a detailed study of heavy
quark phenomena
associated with nucleon structure, such as the charm and bottom structure
functions over
the full range of $x_{bj}$.  The EMC experiment measured a charm
structure
function at large $x_{bj}$ and $Q^2$ much larger than expected from photon gluon
fusion, indicating an intrinsic charm content\cite{IC}
of the proton with probability
$P_{c\bar c} \cong
0.6 \pm 0.3\%$.\cite{HSV}  The intrinsic heavy quark wavefunction is
maximal at equal
rapidity;
\ie\ light-cone momentum fractions
\begin{equation}
k^+_i/p^+ \cong x_{\perp_i}/\sum_j x_{\perp_j} \;
\end{equation}
predicting charm and bottom distributions at large $x_{bj}$ and leading heavy
hadrons at large $x_F$ in the proton fragmentation region.  These
novel features would be well-studied in a collider such as EPIC.

\item
The full range of Compton processes would be accessible, including
exclusive virtual
Compton scattering and Compton/Bethe-Heitler interference $\gamma^* p
\rightarrow
\gamma p^\prime$ and $\gamma^* p \rightarrow \gamma^* p^\prime$;\cite
{Brodsky:1972vv,Brodsky:1973hm}
and inclusive
deep inelastic Compton scattering $\gamma p \rightarrow \gamma X$ and $\gamma p
\rightarrow \gamma^* p$ which can be used to measure new quark
distributions and sum
rules weighted by the quark charge $e^4_q$ or $e^3_q$.\cite{Brodsky:1972yx}

\end{enumerate}

The common denominator in all of these QCD studies is the hadron
light-cone wavefunction; the boost-invariant amplitude which represents
the hadron in terms of its quark and gluon quanta. In the next
section I will review some of the universal features of light-cone
wavefunctions, and in the following sections, I will  review in more
detail specific EPIC topics.

\section{QCD Phenomena at EPIC and the Light-Cone Wavefunctions of Hadrons}

In a relativistic collision, the incident hadron projectile
presents itself as an ensemble of coherent states containing various numbers
of quark and gluon quanta.  Thus when the electron in an electron-proton
collider crosses a
proton at fixed ``light-cone" time
$\tau = t+z/c= x^0 + x^z$, it
encounters a baryonic state with a given number of quarks, anti-quarks, and
gluons in flight with $n_q - n_{\bar q} = 3$.  The natural framework for
describing these hadronic components in QCD at the amplitude level is the
light-cone Fock
representation obtained by quantizing the theory at fixed
$\tau$.\cite{PinskyPauli}
 For example, the proton state has the Fock expansion
\begin{eqnarray}
\ket p &=& \sum_n \VEV{n\,|\,p}\, \ket n \nonumber \\
&=& \psi^{(\Lambda)}_{3q/p} (x_i,\vec k_{\perp i},\lambda_i)\,
\ket{uud} \\[1ex]
&&+ \psi^{(\Lambda)}_{3qg/p}(x_i,\vec k_{\perp i},\lambda_i)\,
\ket{uudg} + \cdots \nonumber
\label{eq:b}
\end{eqnarray}
representing the expansion of the exact QCD eigenstate on a non-interacting
quark and gluon basis.  The probability amplitude
for each such
$n$-particle state of on-mass shell quarks and gluons in a hadron is given by a
light-cone Fock state wavefunction
$\psi_{n/H}(x_i,\vec k_{\perp i},\lambda_i)$, where the constituents have
longitudinal light-cone momentum fractions
\begin{equation}
x_i = \frac{k^+_i}{p^+} = \frac{k^0+k^z_i}{p^0+p^z}\ ,\quad \sum^n_{i=1} x_i= 1
\ ,
\label{eq:c}
\end{equation}
relative transverse momentum
\begin{equation}
\vec k_{\perp i} \ , \quad \sum^n_{i=1}\vec k_{\perp i} = \vec 0_\perp \ ,
\label{eq:d}
\end{equation}
and helicities $\lambda_i.$ The effective lifetime of each configuration
in the laboratory frame is ${2 P_{lab}/{\M}_n^2- M_p^2} $ where
\begin{equation}
\M^2_n = \sum^n_{i=1} \frac{k^2_\perp + m^2}{x} < \Lambda^2 \label{eq:a}
\end{equation} is the off-shell invariant mass and $\Lambda$ is a global
ultraviolet regulator.  The form of the $\psi^{(\Lambda)}_{n/H}(x_i,
\vec k_{\perp i},\Lambda_c)$ is invariant under longitudinal boosts; \ie,\ the
light-cone wavefunctions expressed in the relative coordinates $x_i$ and
$k_{\perp i}$ are independent of the total momentum
$P^+$,
$\vec P_\perp$ of the hadron.

Thus the interactions of the proton reflects an average over the interactions
of its fluctuating states.  For example, a valence state with small impact
separation, and
thus a small color dipole moment, would be expected to interact weakly in a
hadronic or nuclear target reflecting its color transparency.  The nucleus
thus filters differentially different hadron
components.\cite{Bertsch,MillerFrankfurtStrikman} The ensemble
\{$\psi_{n/H}$\} of such light-cone Fock
wavefunctions is a key concept for hadronic physics, providing a conceptual
basis for representing physical hadrons (and also nuclei) in terms of their
fundamental quark and gluon degrees of freedom.  Given the
$\psi^{(\Lambda)}_{n/H},$ we can construct any spacelike electromagnetic or
electroweak form factor from the diagonal overlap of the LC
wavefunctions.\cite{BD}  In the case of semileptonic decays of heavy
mesons, one also obtains important contributions from LC Fock states
with $\Delta n = 2$.\cite{Brodsky:1998hn}   Similarly, the matrix
elements of the currents that define quark and gluon structure functions
can be computed from the integrated squares of the LC
wavefunctions.\cite{BL}

It is thus important not only to compute the
spectrum of hadrons and gluonic states, but also to
determine the wavefunction of each QCD bound state in terms of its
fundamental quark and gluon degrees of freedom.  If we could obtain such
nonperturbative solutions of QCD, then we would be able to compute from
first principles  the
quark and gluon structure functions and distribution amplitudes which control
hard-scattering inclusive and exclusive reactions as well as calculate the
matrix elements of currents which underlie electroweak form factors and the weak
decay amplitudes of the light and heavy hadrons.  The light-cone wavefunctions
also determine the multi-parton correlations which control the distribution of
particles in the proton fragmentation region as well as dynamical higher twist
effects.  Thus one can analyze not only the deep inelastic structure
functions but also the fragmentation of the proton spectator system.
Knowledge of hadron wavefunctions would also open a window to a deeper
understanding of the physics of QCD at the amplitude level, illuminating exotic
effects of the theory such as color transparency, intrinsic heavy quark
effects, hidden color, diffractive processes, and the QCD van der Waals
interactions.

Solving a quantum field theory such as QCD is clearly not easy.  However, highly
non-trivial, one-space one-time relativistic quantum field theories which
mimic many of the features of QCD, have already been completely solved
using light-cone
Hamiltonian methods.\cite{PinskyPauli} Virtually any
(1+1) quantum field theory can be solved using the method of Discretized
Light-Cone-Quantization (DLCQ).\cite{DLCQ,Schlad} In DLCQ, the Hamiltonian
$H_{LC}$, which can be constructed from the Lagrangian using light-cone time
quantization, is completely diagonalized, in analogy to Heisenberg's solution of
the eigenvalue problem in quantum mechanics.  The quantum field theory
problem is
rendered discrete by imposing periodic or anti-periodic boundary
conditions.  The eigenvalues and eigensolutions of collinear QCD then give the
complete spectrum of hadrons, nuclei, and gluonium and their respective
light-cone wavefunctions.

The existence of an exact formalism
provides a basis for systematic approximations.  For example, one can
analyze exclusive processes
which involve hard internal momentum transfer using a
perturbative QCD formalism patterned after the analysis of form factors at
large momentum transfer.\cite{BL} The hard-scattering analysis proceeds by
writing each hadronic wavefunction as a sum of soft and hard contributions
$\psi_n = \psi^{{\rm soft}}_n (\M^2_n < \Lambda^2) + \psi^{{\rm hard}}_n
(\M^2_n >\Lambda^2) ,$
where $\M^2_n $ is the invariant mass of the partons in the $n$-particle
Fock state and
$\Lambda$ is the separation scale.
The high internal momentum contributions to the wavefunction $\psi^{{\rm
hard}}_n $ can be calculated systematically from QCD perturbation theory
by iterating the gluon exchange kernel.  The contributions from high
momentum transfer
amplitude can then be written as a convolution of a hard scattering
quark-gluon scattering amplitude $T_H$ with the distribution
amplitudes $\phi(x_i,\Lambda)$, the valence wavefunctions obtained by
integrating the
constituent momenta up to the separation scale
${\cal M}_n < \Lambda < Q$.  This is the basis for the perturbative hard
scattering analyses.\cite{BHS,Sz,BALL,BABR}
In the exact analysis, one can
identify the hard PQCD contribution as well as the soft contribution from
the convolution of the light-cone wavefunctions.  Furthermore, the hard
scattering contribution can be systematically improved.  For example, off-shell
effects can be retained in the evaluation of
$T_H$ by utilizing the exact light-cone energy denominators.

More generally, hard exclusive hadronic amplitudes such as quarkonium
decay, heavy hadron decay,  and scattering amplitudes such as deeply
virtual Compton
scattering can be written as the convolution of the light-cone Fock state
wavefunctions with
quark-gluon matrix elements
\cite{BL}
\begin{eqnarray}
\M_{\rm Hadron} &=& \prod_H \sum_n \int
\prod^{n}_{i=1} d^2k_\perp \prod^{n}_{i=1}dx\, \delta
\left(1-\sum^n_{i=1}x_i\right)\, \delta
\left(\sum^n_{i=1} \vec k_{\perp i}\right) \nonumber \\[2ex]
&& \times \psi^{(\Lambda)}_{n/H} (x_i,\vec k_{\perp i},\Lambda_i)\,
T_H^{(\Lambda)} \ .
\label{eq:e}
\end{eqnarray}
Here $T_H^{(\Lambda)}$ is the underlying quark-gluon
subprocess scattering amplitude, where the (incident or final) hadrons are
replaced by quarks and gluons with momenta $x_ip^+$, $x_i\vec
p_{\perp}+\vec k_{\perp i}$ and invariant mass above the
separation scale $\M^2_n > \Lambda^2$.
The essential part of the wavefunction is the hadronic distribution amplitudes,
\cite{BL} defined as the integral over transverse momenta of the valence (lowest
particle number) Fock wavefunction; \eg\ for the pion
\begin{equation}
\phi_\pi (x_i,Q) \equiv \int d^2k_\perp\, \psi^{(Q)}_{q\bar q/\pi}
(x_i, \vec k_{\perp i},\lambda)
\label{eq:f}
\end{equation}
where the global cutoff $\Lambda$ is identified with the
resolution $Q$.  The distribution amplitude controls leading-twist exclusive
amplitudes at high momentum transfer, and it can be related to the
gauge-invariant Bethe-Salpeter wavefunction at equal light-cone time
$\tau = x^+$.  The $\log Q$ evolution of the hadron distribution amplitudes
$\phi_H (x_i,Q)$ can be derived from the
perturbatively-computable tail of the valence light-cone wavefunction in the
high transverse momentum regime.\cite{BL} In general the LC ultraviolet
regulators provide a factorization scheme for elastic and inelastic
scattering, separating the hard dynamical contributions with invariant mass
squared $\M^2 > \Lambda^2_{\rm global}$ from the soft physics with
$\M^2 \le \Lambda^2_{\rm global}$ which is incorporated in the
nonperturbative LC wavefunctions.  The DGLAP evolution of quark and gluon
distributions can also be derived by computing the variation of the Fock
expansion with respect to $\Lambda^2$.\cite{BL}

Given the solution
for the hadronic wavefunctions $\psi^{(\Lambda)}_n$ with $\M^2_n <
\Lambda^2$, one can construct the wavefunction in the hard regime with
$\M^2_n > \Lambda^2$ using projection operator techniques.\cite{BL} The
construction can be done perturbatively in QCD since only high invariant mass,
far off-shell matrix elements are involved.  One can use this method to
derive the physical properties of the LC wavefunctions and their matrix elements
at high invariant mass.  Since $\M^2_n = \sum^n_{i=1}
\left(\frac{k^2_\perp+m^2}{x}\right)_i $, this method also allows the derivation
of the asymptotic behavior of light-cone wavefunctions at large $k_\perp$, which
in turn leads to predictions for the fall-off of form factors and other
exclusive
matrix elements at large momentum transfer, such as the quark counting rules
for predicting the nominal power-law fall-off of two-body scattering amplitudes
at fixed
$\theta_{cm}.$\cite{BL} The phenomenological successes of these rules
can be understood within QCD if the coupling
$\alpha_V(Q)$ freezes in a range of relatively
small momentum transfer.\cite{BJPR}

\section{Measurement of Light-Cone Wavefunctions via Diffractive Dissociation}

Diffractive multi-jet production in heavy
nuclei provides a novel way to measure the shape of the LC Fock
state wavefunctions and test color transparency.  For example, consider the
reaction\cite{Bertsch,MillerFrankfurtStrikman}
$\pi A \rightarrow {\rm Jet}_1 + {\rm Jet}_2 + A^\prime$
at high energy where the nucleus $A^\prime$ is left intact in its ground
state.  The transverse momenta of the jets have to balance so that
$
\vec k_{\perp i} + \vec k_{\perp 2} = \vec q_\perp < \R^{-1}_A \ ,
$
and the light-cone longitudinal momentum fractions have to add to
$x_1+x_2 \sim 1$ so that $\Delta p_L < R^{-1}_A$.  The process can
then occur coherently in the nucleus.  Because of color transparency,  i.e.,
the cancelation of color interactions in a small-size color-singlet
hadron,  the valence wavefunction of the pion with small impact
separation will penetrate the nucleus with minimal interactions,
diffracting into jet pairs.  \cite{Bertsch}
The $x_1=x$, $x_2=1-x$ dependence of
the di-jet distributions will thus reflect the shape of the pion distribution
amplitude; the $\vec k_{\perp 1}- \vec k_{\perp 2}$
relative transverse momenta of the jets also gives key information on
the underlying shape of the valence pion wavefunction.
The QCD analysis\cite{MillerFrankfurtStrikman}can be
confirmed by the observation that the diffractive nuclear amplitude
extrapolated to $t = 0$ is linear in nuclear number $A$, as predicted by
QCD color
transparency.  The integrated diffractive rate should scale
approximately as
$A^2/R^2_A \sim A^{4/3}$.  A diffractive dissociation experiment of this
type, E791,  is now in progress at Fermilab using 500 GeV incident pions
on nuclear targets.\cite{E791} The preliminary results from E791 appear
to be consistent with color transparency.  The momentum fraction
distribution of the jets is consistent with a valence light-cone
wavefunction of the pion consistent with the shape of the asymptotic
distribution amplitude, the asymptotic solution \cite{BL}
$\phi^{\rm asympt}_\pi (x) =
\sqrt 3 f_\pi x(1-x)$ to the perturbative QCD evolution
equation.  Data from CLEO for the
$\gamma
\gamma^* \rightarrow \pi^0$ transition form factor also favor a form for
the pion distribution amplitude close to this form.\cite{Kroll,Rad,BJPR}
Conversely, one can use incident real and virtual photons:
$ \gamma^* A \rightarrow {\rm Jet}_1 + {\rm Jet}_2 + A^\prime $ to
 at EPIC confirm the shape of the calculable light-cone wavefunction for
transversely-polarized and longitudinally-polarized virtual photons.  Such
experiments will open up a remarkable, direct window on the amplitude
structure of hadrons at short distances. Most interesting, one
can use the EPIC electron beam to diffractively dissociate the proton beam
into three high $p_T$ jets in the proton fragmentation region
$ep \rightarrow
e^\prime {\cal J}_1 {\cal J}_2 {\cal J}_3$ reflecting the high transverse
momentum
components of the valance proton
wavefunction.\cite{MillerFrankfurtStrikman}

\section{Other Applications of Light-Cone Quantization to EPIC QCD
Phenomenology}

{\it Diffractive vector meson photoproduction.} The
light-cone Fock wavefunction representation of hadronic amplitudes
provides a simple eikonal analysis of diffractive high energy processes, such as
$\gamma^*(Q^2) p \to \rho p$, in terms of the virtual photon and the vector
meson Fock state light-cone wavefunctions convoluted with the $g p \to g p$
near-forward matrix element.\cite{BGMFS} One can easily show that only small
transverse size $b_\perp \sim 1/Q$ of the vector meson distribution
amplitude is involved.  The hadronic interactions are minimal, and thus the
$\gamma^*(Q^2) N \to
\rho N$ reaction can occur coherently throughout a nuclear target in reactions
without absorption or shadowing.  The $\gamma^* A \to V A$ process thus
provides a natural framework for testing QCD color transparency.\cite{BM}
Evidence for
color transparency in such reactions has been found by Fermilab experiment
E665.

{\it Regge behavior of structure functions.} The light-cone wavefunctions
$\psi_{n/H}$ of a hadron are not independent of each other, but rather are
coupled via the equations of motion.  Antonuccio, Dalley and I
\cite{ABD} have used the constraint of finite ``mechanical'' kinetic energy to
derive ``ladder relations" which interrelate the light-cone wavefunctions of
states differing by one or two gluons.  We then use these relations to
derive the
Regge behavior of both the polarized and unpolarized structure functions at $x
\rightarrow 0$, extending Mueller's derivation of the BFKL hard
QCD pomeron from the properties of heavy quarkonium light-cone wavefunctions at
large $N_C$ QCD.\cite{Mueller}

{\it Structure functions at large $x_{bj}$.} The behavior of structure functions
where one quark has the entire momentum requires the knowledge of LC
wavefunctions
with $x \rightarrow 1$ for the struck quark and $x \rightarrow 0$ for the
spectators.  This is a highly off-shell configuration, and thus one can
rigorously derive quark-counting and helicity-retention rules for the
power-law behavior of
the polarized and unpolarized quark and gluon distributions in the $x
\rightarrow 1$ endpoint domain.\cite{BL}  It is interesting to note that the
evolution of structure functions is minimal in this domain because the
struck quark is highly virtual as $x\rightarrow 1$; \ie\ the starting point
$Q^2_0$ for evolution
cannot be held fixed, but must be larger than a scale of order
$(m^2 + k^2_\perp)/(1-x)$.\cite{BL,Dmuller}

{\it Intrinsic gluon and heavy quarks.}
The main features of the heavy sea quark-pair contributions of the Fock
state expansion of light hadrons can also be derived from perturbative QCD,
since $\M^2_n$ grows with
$m^2_Q$.  One identifies two contributions to the heavy quark sea, the
``extrinsic'' contributions which correspond to ordinary gluon splitting, and
the ``intrinsic" sea which is multi-connected via gluons to the valence quarks.
The intrinsic sea is thus sensitive to the hadronic bound state
structure.\cite{IC} The maximal contribution of the
intrinsic heavy quark occurs at $x_Q \simeq {m_{\perp Q}/ \sum_i m_\perp i}$
where $m_\perp = \sqrt{m^2+k^2_\perp}$;
\ie\ at large $x_Q$, since this minimizes the invariant mass $\M^2_n$.  The
measurements of the charm structure function by the EMC experiment are
consistent with intrinsic charm at large $x$ in the nucleon with a
probability of order $0.6 \pm 0.3 \% $.\cite{HSV} Similarly, one can
distinguish intrinsic
gluons which are associated with multi-quark interactions and extrinsic gluon
contributions associated with quark substructure.\cite{BS} One can also use this
framework to isolate the physics of the anomaly contribution to the Ellis-Jaffe
sum rule.

{\it Materialization of far-off-shell configurations.}
In a high energy hadronic collisions, the highly-virtual states of a hadron
can be
materialized into physical hadrons simply by the soft interaction of any of the
constituents.\cite{BHMT} Thus a proton state with intrinsic charm $\ket{ u
u d \bar c c}$ can be materialized, producing a $J/\psi$ at large $x_F$,  by the
interaction of the electron with the proton beam.

{\it Comover phenomena.}
Light-cone wavefunctions describe not only the partons that interact in a hard
subprocess but also the associated partons freed from the projectile.  The
projectile partons which are comoving (\ie, which have similar rapidity) with
final state quarks and gluons can interact strongly producing (a) leading
particle effects, such as those seen in open charm hadroproduction; (b)
suppression of quarkonium\cite{BrodskyMueller} in favor of open heavy hadron
production, as seen in the E772 experiment; (c) changes in color configurations
and selection rules in quarkonium hadroproduction, as has been emphasized by
Hoyer and Peigne.\cite{HoyerPeigne} All of these effects violate the
usual ideas of factorization for inclusive reactions.  More than
one parton from the
projectile can enter the hard subprocess, producing dynamical higher twist
contributions, as seen for example in
Drell-Yan experiments.\cite{BrodskyBerger,Brandenburg}

{\it Asymmetric sea.}\
In conventional studies of the ``sea'' quark distributions, it is
usually assumed that, aside from the effects due to antisymmetrization
with valence quarks, the quark and antiquark sea contributions have the
same momentum and helicity distributions.  However, the ansatz of
identical quark and anti-quark sea contributions has never been
justified, either theoretically or empirically.  Obviously the sea
distributions which arise directly from gluon splitting in leading
twist are necessarily CP-invariant; \ie,\ they are symmetric under
quark and antiquark interchange.  However, the initial distributions
which provide the boundary conditions for QCD evolution need not be
symmetric since the nucleon state is itself not CP-invariant.  Only the
global quantum numbers of the nucleon must be conserved.  The intrinsic
sources of strange (and charm) quarks reflect the wavefunction
structure of the bound state itself; accordingly, such distributions
would not be expected to be CP symmetric.\cite{BrodskyMaSigBur,MA} Thus the
strange/anti-strange asymmetry of nucleon structure functions provides
a direct window into the quantum bound-state structure of hadronic
wavefunctions.

Quark
and antiquark asymmetry is also implied by a
light-cone meson-baryon fluctuation model of intrinsic $q\bar q$
pairs.\cite{BrodskyMaSigBur} The most important
fluctuations are those closest to the energy shell with minimal
invariant mass.  For example, the coupling of a proton to a virtual
$K^+ \Lambda$ pair provides a specific source of intrinsic strange
quarks and antiquarks in the proton.  Since the $s$ and $\bar s$
quarks appear in different configurations in the lowest-lying hadronic
pair states, their helicity and momentum distributions are distinct.
Such fluctuations are necessarily part of any quantum-mechanical
description of the hadronic bound state in QCD and have also been
incorporated into the cloudy bag model and Skyrme
solutions to chiral theories.  Ma and I \cite{MA} have utilized a
boost-invariant light-cone Fock state description of the hadron
wavefunction which emphasizes multi-parton configurations of minimal
invariant mass.  We find that such fluctuations predict a striking sea
quark and antiquark asymmetry in the corresponding momentum and
helicity distributions in the nucleon structure functions.  In
particular, the strange and anti-strange distributions in the nucleon
generally have completely different momentum and spin characteristics.
For example, the model predicts that the intrinsic $d$ and $s$ quarks
in the proton sea are negatively polarized, whereas the intrinsic $\bar
d$ and $\bar s$ antiquarks provide zero contributions to the proton
spin.  We also predict that the intrinsic charm and anticharm helicity
and momentum distributions are not strictly identical.  The above
picture of quark and antiquark asymmetry in the momentum and helicity
distributions of the nucleon sea quarks has support from a number of
experimental observations, and we suggest processes to test and measure
this quark and antiquark asymmetry in the nucleon sea.

In addition, one could identify dynamical higher twist processes
where the
electron recoils against two quarks versus one quark by studying the
pattern of jet
and dijet production in the current and proton beam
fragmentation region.

{\it Hidden Color }
The deuteron form factor at high $Q^2$ is sensitive to wavefunction
configurations where all six quarks overlap within an impact
separation $b_{\perp i} < {\cal O} (1/Q);$ the leading power-law
fall off predicted by QCD is $F_d(Q^2) = f(\alpha_s(Q^2))/(Q^2)^5$,
where, asymptotically, $f(\alpha_s(Q^2)) \propto
\alpha_s(Q^2)^{5+2\gamma}$. \break \cite{bc76,bjl83}
In general, the six-quark wavefunction of a deuteron
is a mixture of five different color-singlet states.  The dominant
color configuration at large distances corresponds to the usual
proton-neutron bound state.  However at small impact space
separation, all five Fock color-singlet components eventually
acquire equal weight, \ie, the deuteron wavefunction evolves to
80\%\ ``hidden color.'' The relatively large normalization of the
deuteron form factor observed at large $Q^2$ points to sizable
hidden color contributions.\cite{fhzxx}

{\it Spin-Spin Correlations in Nucleon-Nucleon
Scattering and the Charm \hfill\break Threshold }
One of the most striking anomalies in elastic proton-proton
scattering is the large spin correlation $A_{NN}$ observed at large
angles.\cite{krisch92} At $\sqrt s \simeq 5 $ GeV, the rate for
scattering with incident proton spins parallel and normal to the
scattering plane is four times larger than that for scattering with
anti-parallel polarization.  This strong polarization correlation can
be attributed to the onset of charm production in the intermediate
state at this energy.\cite{bdt88} The intermediate state $\vert u
u d u u d c \bar c \rangle$ has odd intrinsic parity and couples to
the $J=S=1$ initial state, thus strongly enhancing scattering when
the incident projectile and target protons have their spins parallel
and normal to the scattering plane.  The charm threshold can also
explain the anomalous change in color transparency observed at the
same energy in quasi-elastic $ p p$ scattering.  A crucial test is
the observation of open charm production near threshold with a
cross
section of order of $1 \mu$b.
One can also expect similar strong spin-spin correlations at the
threshold for charm and bottom production in photon-proton collisions at EPIC.

\section{Semi-Exclusive Processes:  New Probes of Hadron Structure at
EPIC}

A new class of hard ``semi-exclusive''
processes of the form $A+B \to C + Y$, have been proposed as new probes of
QCD.\cite{BB,acw,Brodsky:1998sr} These processes are characterized
by a large momentum transfer $t= (p_A-p_C)^2$ and a large rapidity gap between
the final state particle $C$ and the inclusive system $Y$.
Here $A, B$
and $C$ can be hadrons or (real or virtual) photons.  The cross
sections for such processes factorize in terms of the distribution
amplitudes of $A$ and $C$ and the parton distributions in the target
$B$.  Because of this factorization semi-exclusive reactions provide a
novel array of generalized currents, which not only give insight into
the dynamics of hard scattering QCD processes, but also allow
experimental access to new combinations of the universal quark and
gluon distributions.  These are ideal processes to study at EPIC.

QCD scattering amplitudes for
deeply virtual exclusive processes such as  Compton scattering $\gamma^* p
\to \gamma p$ and meson production $\gamma^* p \to M p$ factorize
into a hard subprocess and soft universal hadronic matrix
elements.  \cite{JiRad,CFS,BGMFS}
For example, consider exclusive meson
electroproduction such as $e p \to e \pi^+ n$ (Fig.~1a).  Here one
takes (as in DIS) the Bjorken limit of large photon virtuality, with
$x_B = Q^2/(2 m_p \nu)$ fixed, while the momentum transfer $t =
(p_p-p_n)^2$ remains small.  These processes involve `skewed' parton
distributions, which are generalizations of the usual parton
distributions measured in DIS.  The skewed distribution in Fig.~1a
describes the emission of a $u$-quark from the proton target together
with the formation of the final neutron from the $d$-quark and the
proton remnants.  As the subenergy $\hat s$ of the scattering process
$\gamma^* u \to \pi^+ d$ is not fixed, the amplitude involves an
integral over the $u$-quark momentum fraction $x$.
\begin{figure*}
\begin{center}
  \leavevmode
  \epsfxsize=3.5in
 \epsfbox{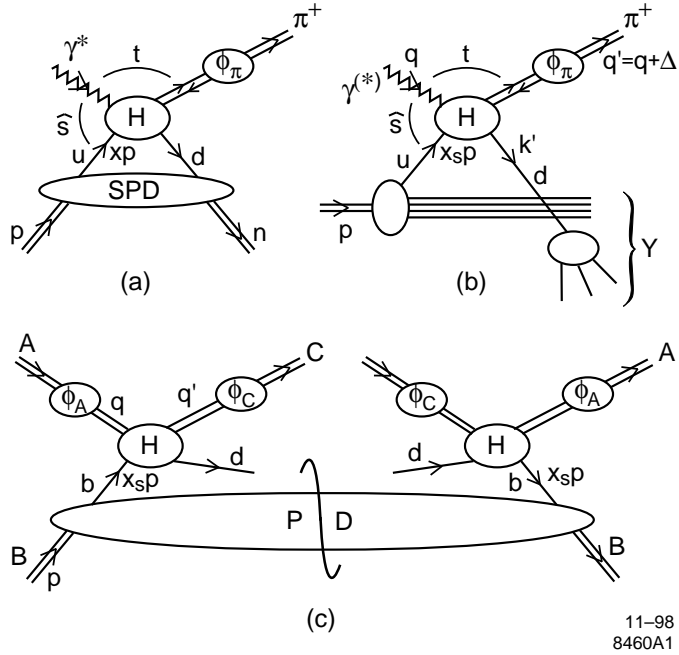}
\end{center}
\caption{{} (a): Factorization of $\gamma^* p \to \pi^+ n$ into a
 skewed parton distribution (SPD), a hard scattering $H$ and the pion
 distribution amplitude $\phi_\pi$.  (b): Semi-exclusive process
 $\gamma^{(*)} p \to \pi^+ Y$.  The $d$-quark produced in the hard
 scattering $H$ hadronizes independently of the spectator partons in
 the proton.  (c): Diagram for the cross section of a generic
 semi-exclusive process.  It involves a hard scattering $H$,
 distribution amplitudes $\phi_A$ and $\phi_C$ and a parton
 distribution (PD) in the target $B$.}
\end{figure*}

An essential condition for the factorization of the deeply virtual
meson production amplitude of Fig.\ 1a is the existence of a large
rapidity gap between the produced meson and the neutron.  This
factorization remains valid if the
neutron is replaced with a hadronic system $Y$ of invariant mass $M_Y^2 \ll
W^2$, where $W$ is
the c.m.\ energy of the $\gamma^* p$ process.
For $M_Y^2 \gg m_p^2$ the momentum $k'$ of the $d$-quark in Fig.~1b is
large with respect to the proton remnants, and hence it forms a jet.
This jet hadronizes independently of the other particles in the final
state if it is not in the direction of the meson, \ie, if the meson
has a large transverse momentum $q'_\perp = \Delta^{\phantom .}_\perp$
with respect to the photon direction in the $\gamma^* p$ c.m.  Then the
cross section for an inclusive system $Y$ can be calculated as in DIS,
by treating the $d$-quark as a final state particle.

The large $\dpt$ furthermore allows only transversally compact
configurations of the projectile $A$ to couple to the hard subprocess
$H$ of Fig.~1b, as it does in exclusive processes.  \cite{BL} Hence the
above discussion applies not only to incoming virtual photons at large
$Q^2$, but also to real photons $(Q^2=0)$ and in fact to any hadron
projectile.

Let us then consider the general process $A+B\to C+Y$, where $B$ and
$C$ are hadrons or real photons, while the projectile $A$ can also be
a virtual photon.  In the semi-exclusive kinematic limit
$\lqcd^2,\, M_B^2,\, M_C^2 \ll
M_Y^2,\, \dpt^2 \ll W^2$
we have a large rapidity gap
$|y_C - y_d| = \log \frac{W^2}{\dpt^2 + M_Y^2}$
between $C$ and the parton $d$ produced in the hard scattering (see
Fig.~1c). The cross section then factorizes into the form
\begin{eqnarray}
\lefteqn{ \frac{d\sigma}{dt\,dx_S}(A+B\to C+Y) }
 \hspace{4em} \nonumber \\
&=& \sum_{b} f_{b/B}(x_S,\mu^2) \frac{d\sigma}{dt} (A b \to C d)
 \eqcm \label{gencross}
\end{eqnarray}
where $t=(q-q')^2$ and $f_{b/B}(x_S,\mu^2)$ denotes the distribution
of quarks, antiquarks and gluons $b$ in the target $B$.  The momentum
fraction $x_S$ of the struck parton $b$ is fixed by kinematics to the
value
$x_S = \frac{-t}{M_Y^2-t}$
and the factorization scale $\mu^2$ is characteristic of the hard
subprocess $A b \to C d$.

It is conceptually helpful to regard the hard scattering amplitude $H$
in Fig.~1c as a generalized current of momentum $q-q'=p_A - p_C$,
which interacts with the target parton $b$.  For $A= \gamma^*$ we
obtain a close analogy to standard DIS when particle $C$ is removed.
With $q' \to 0$ we thus find $-t \to Q^2$, $M_Y^2 \to W^2$, and see
that $x_S$ goes over into $x_B = Q^2 /(W^2 + Q^2)$.  The
possibility to control the value of $q'$ (and hence the momentum
fraction $x_S$ of the struck parton) as well as the quantum numbers of
particles $A$ and $C$ should make semi-exclusive processes a versatile
tool for studying hadron structure.  The cross section further depends
on the distribution amplitudes $\phi_A$, $\phi_C$ (\cf\ Fig.~1c),
allowing new ways of measuring these quantities.  The use of this new
current requires a sufficiently high c.m.\ energy, since we need to have at
least one intermediate large scale.  The possibility of creating effective
currents using
processes similar to the ones we discuss here was considered already
before the advent of QCD.  \cite{BB}

In the case of photoproduction one finds in the limit
with $Q^2=0$ that the $\gamma u \to \pi^+ d$ subprocess cross section
is \cite{acw}
\begin{eqnarray}
\lefteqn{\frac{d\sigma}{dt}(\gamma u \to \pi^+ d) =
          \frac{128\pi^2}{27}\, \alpha\as^2\,
          \frac{(e_u-e_d)^2}{{\hat s}^2 (-t)} } \hspace{3em} \inn \\
&\times& \left\{\left[\int_0^1 dz \frac{\phi_\pi(z)}{z} \right]^2 +
\left[\int_0^1 dz \frac{\phi_\pi(z)}{1-z} \right]^2 \right\} \eqcm
  \label{photocross}
\end{eqnarray}
where $\int dz\, \phi_\pi(z) = f_\pi /
\sqrt{12}$ with $f_\pi = 93$ MeV.  The result for the physical
process $\gamma p \to \pi^+ Y$ is then,
\begin{eqnarray}
\lefteqn{ \frac{d\sigma}{dt\,dx_S}(\gamma p \to \pi^+ Y) }
 \hspace{2em} \inn \\
&=& \left[u(x_S,-t) + \bar d(x_S,-t) \right]
 \frac{d\sigma}{dt}(\gamma u \to \pi^+ d) \label{fullphoto}
\end{eqnarray}
with the notation $q(x_S,-t) = f_{q/p}(x_S,-t)$. The pion distribution
amplitude $\phi_\pi(z)$ enters in
 precisely the same way as it does in the pion transition form factor
 for $\gamma^* \gamma \to \pi^0$ \cite{BL,acw}
\begin{equation}
F_{\pi\gamma}(Q^2) = {\sqrt{48}\,
(e_u^2- e_d^2) \over Q^2} \int^1_0 dz {\phi_\pi(z)\over z} \eqpt
\label{fpigam}
\end{equation}

There are several
interesting aspects of this result:
Both the target $u$- and $\bar d$-quark contributions are
 weighted by the {\em total charge} $e_u-e_d = +1$ of the produced
 $\pi^+$.  An analogous formula holds of course if the $\pi^+$ is
 replaced with another pseudoscalar.  For neutral meson production,
 $\gamma p \to M^0 Y$ with $M^0=\pi^0$, $K^0$, $\eta$, \ldots, the
 expression (\ref{photocross}) \emph{vanishes}; more exactly one
 finds that the cross section is suppressed by $(-t /\hat{s})^2$
 compared with the charged meson case.  \cite{acw}
Note that the cross section has a power-law behavior, $d
 \sigma/ d t \propto 1 /\hat{s}^3$ at fixed $t /\hat{s}$.  This is the
 basic signature that the amplitude factorizes into a meson
 distribution amplitude and a hard scattering subprocess.
At fixed $t$ the expression (\ref{photocross}) goes like $1
 /\hat s^2$, which is characteristic of two spin 1/2 quark exchanges
 in the $t$-channel.  Notice that with $\lqcd^2 \ll -t \ll \hat{s}$
 the hard scattering takes place in the perturbative Regge regime: if
 a deviation from this $\hat{s}$-behavior were to be observed
 experimentally it would indicate that the quark exchange reggeizes,
 \ie, that contributions from higher order ladder diagrams are
 important. At
large $-t$, non-singlet Regge trajectories asymptote to 0 or negative
integers, reflecting their quark pair composition.\cite{BTT}

In the semi-exclusive limit with $Q^2 \sim W^2$, and finite $x_B$, the
semi-exclusive electroproduction cross section becomes
\begin{eqnarray}
\lefteqn{ \frac{d\sigma(ep\to e\pi^+Y)}{dQ^2\,dx_B\,dt\,dx_S} =
\frac{\alpha}{\pi} \frac{1-y}{Q^2 x_B}\,
\frac{512\pi^2}{27}\, \alpha\as^2
\frac{x_B}{\hat{s}\, Q^4 x_S}} \hspace{0.5em} \nonumber\\
&&
\times \left[\int_0^1 dz \frac{\phi_\pi(z)}{z} \right]^2
\left\{ u(x_S) \left[e_u+
\Big(1-\frac{x_B}{x_S}\Big)\, e_d \right]^2 \right.
\nonumber \\
&& \hspace{7.7em} + \left.
\bar d(x_S) \left[e_d+
\Big(1-\frac{x_B}{x_S}\Big)\, e_u \right]^2 \right\}
\label{elcross}
\end{eqnarray}
where $y=\nu/E_e$ is the momentum fraction of the projectile electron
carried by the virtual photon, and we have used again $\phi_\pi(z)=
\phi_\pi(1-z)$.  The semi-exclusive cross section in \eq{elcross} corresponds to
longitudinal photon exchange.  The contribution from transverse
 photons is suppressed, as in the exclusive case $\gamma^* p \to Mp$
 at large $Q^2$ and small $-t$.  \cite{CFS}

The systematic comparisons of
semi-exclusive photoproduction of various particles at a facility such as EPIC
thus can give useful
information on parton distributions and distribution amplitudes.  The
hard subprocess (\ref{photocross}) cancels in the ratio of physical
cross sections (\ref{fullphoto}) for $\pi^+$ and $\pi^-$.  Hence
$d\sigma(\pi^+)/d\sigma(\pi^-)$ directly measures the $(u+\bar
d)/(d+\bar u)$ parton distribution ratio.  Similarly, the
$d\sigma(K^+)/d\sigma(K^-)$ ratio measures the strange quark content
of the target without uncertainties due, \eg, to fragmentation
functions. Conversely, the parton distributions drop out in the ratio
$d\sigma(\rho_L^+)/d\sigma(\pi^+)$ of longitudinally polarized $\rho$
mesons to pions, allowing a comparison of their distribution
amplitudes.  Since the normalization of both $\phi_\rho$ and $\phi_\pi$
is fixed by the leptonic decay widths such a comparison can reveal
differences in their $z$-dependence.  In the intermediate $Q^2$-range
the relative size of $Q^2$ and $t$ can furthermore be
tuned to change the dependence of the hard subprocess on $z$ and thus obtain
further information
on the shape of
$\phi(z)$.

{\em Spin and transversity distributions in Semi-Exclusive Reactions.}
The polarization of the
target $B$ can naturally be incorporated in this framework.  A
longitudinally polarized target selects the usual spin-dependent
parton distributions $\Delta q(x_S)$.  It also appears possible to
measure the quark transverse spin, or transversity distribution in
photoproduction of $\rho$ mesons on transversely polarized protons.  In
this case only the interference term between transversely and
longitudinally polarized $\rho$ mesons should contribute.

{\em Color transparency in Semi-Exclusive Reactions.} The factorization of
the hard amplitude
$H$ in Fig.\ 1c from the target remnants is a consequence of the high
transverse momentum which selects compact sizes in the projectile $A$
and in the produced particle $C$.  In the case of nuclear targets this
color transparency \cite{CT} implies according to \eq{gencross} that
all target dependence enters via the nuclear parton distribution.  Thus
tests of color transparency can be made even in photoproduction, \eg,
through
\begin{equation}
\gamma A \to \left\{
\begin{array}{c}
\pi^+(\dpt) + Y \\
p(\dpt) + Y
\end{array} \right. \label{ctest}
\end{equation}
in the semi-exclusive kinematic region.
Color transparency has so far been studied mainly in exclusive
processes where the target scatters elastically, such as $\gamma^*A
\to \rho A$,\cite{e665} $pA \to pp+(A-1)$ \cite{BNL} and $\gamma^*A
\to p+(A-1)$. Semi-exclusive processes provide a
possibility to study color transparency at EPIC where the target
dissociates into a heavy inclusive system $Y$.\cite{PH}

\section{Odderon-Pomeron Interference at EPIC}

The existence of odd charge-conjugation,  zero flavor-number exchange
contributions to high energy hadron scattering
amplitudes is a basic prediction of quantum chromodynamics, following
simply from the existence of the color-singlet exchange
of three reggeized gluons in the
$t-$channel.  \cite{kwiencinski_bartels}
In Regge theory, the ``Odderon" contribution is dual to a sum over
$C = P= -1$ gluonium states in the $t$-channel.
\cite{landshoff_nachtmann,lukaszuk_nicolescu} In the case of
reactions which involve high momentum transfer, the deviation of the Regge
intercept of the Odderon trajectory from $\alpha_{\cal O}(t=0) = 1$ can
in principle be computed~\cite{Lipatov2,Braun,Wosieck,Gauron2}
from perturbative QCD in
analogy to the methods used to compute the properties of the hard BFKL
pomeron.\cite{BFKL}

Recently Rathsman, Merino and I \cite{Brodsky:1999mz}
have proposed
an experimental test well suited to EPIC, COMPASS,  and HERA kinematics which
should be able to disentangle the contributions of both the Pomeron and the
Odderon to diffractive production of charmed jets.  By forming a charge
asymmetry in the energy of the charmed jets, we can determine the
relative importance of the Pomeron ($C=+$) and the Odderon ($C=-$)
contributions, and their interference, thus providing a new
experimental test of the separate existence of these two objects.
Since the asymmetry measures the
Odderon amplitude linearly, even a relatively weakly-coupled amplitude
should be visible.

The leading contributions to the  amplitude for diffractive
photoproduction of a charm quark anti-quark pair is given by single Pomeron
exchange~(two reggeized
gluons), and the next term in the Born expansion is given by the exchange
of one Odderon~(three reggeized gluons).
In general the Pomeron and Odderon exchange
amplitudes will interfere.
The contribution of the
interference term to the total cross-section is zero, but it
does contribute to charge-asymmetric rates.  Thus we propose the study of
photoproduction of $c$-$\bar{c}$ pairs and measure the asymmetry in
the energy fractions $z_c$ and $z_{\bar{c}}$.  More generally,
one can use other charge-asymmetric kinematic configurations, as well as
bottom or strange quarks.
The interference term can be isolated by forming
the charge asymmetry,
\begin{equation} \label{eq:asym}
{\cal A}(t,M_X^2,z_c) =
\frac{\displaystyle \frac{d\sigma}{dtdM_X^2dz_c}
                  - \frac{d\sigma}{dtdM_X^2dz_{\bar{c}}} }
{\displaystyle \frac{d\sigma}{dtdM_X^2dz_c}
             + \frac{d\sigma}{dtdM_X^2dz_{\bar{c}}} } \; ,
\end{equation}
The predicted asymmetry has the form
\begin{eqnarray}\label{eq:pred}
&&
{\cal A}(t,M_X^2,z_c) \\
& = & \frac{
{\displaystyle g_{pp^\prime}^{\pom}
g_{pp^\prime}^{\odd}
\left(\frac{s_{\gamma p}}{M_X^2}\right)^{\alpha_{\pom}+\alpha_{\odd} }
\frac{2\sin \left[\frac{\pi}{2}\left(\alpha_{\odd}-\alpha_{\pom}\right)
 \right]}
{\sin{ \frac{\pi\alpha_\pom}{2}} \cos{ \frac{\pi\alpha_\odd}{2}}}
g_{\pom}^{\gamma c\bar{c}}
g_{\odd}^{\gamma c\bar{c}}}
}
{{\displaystyle \left[g_{pp^\prime}^{\pom}
\left(\frac{s_{\gamma p}}{M_X^2}\right)^{\alpha_{\pom} }
g_{\pom}^{\gamma c\bar{c}}
/\sin\frac{\pi\alpha_\pom}{2}
\right]^2 +
\left[g_{pp^\prime}^{\odd}
\left(\frac{s_{\gamma p}}{M_X^2}\right)^{\alpha_{\odd} }
g_{\odd}^{\gamma c\bar{c}}
/\cos\frac{\pi\alpha_\odd}{2}
\right]^2 }} \ .\nonumber
\end{eqnarray}

Thus by observing the charge asymmetry of the charm quark/antiquark energy
fraction ($z_c$)
in diffractive $c\bar{c}$ pair photoproduction, the interference
between the Pomeron and the Odderon exchanges can be isolated and the
ratio to the sum of the Pomeron and the Odderon exchanges measured.
In a simple model for the Pomeron/Odderon coupling to the photon the
asymmetry is predicted to be proportional to $(2z_c-1)/(z_c^2+(1-z_c)^2)$
and the magnitude is of order 15\% (and possibly significantly larger for
diffractive proton dissociation).  Such a test could be performed
by experiments at EPIC, COMPASS,  or HERA measuring the diffractive
production of open charm in photoproduction or electroproduction.
Such measurements could provide the first experimental evidence for the
existence of the Odderon,  as well as the relative
strength of the Odderon and Pomeron couplings.  Most important, the energy
dependence of the asymmetry can be used to determine whether the Odderon
intercept is in fact greater or less than that of the Pomeron.

\section{Summary}
An electron-proton/ion polarized beam collider of medium energy clearly
has strong potential
for future fundamental studies of proton and nuclear structure in QCD.
Only a sample of EPIC topics have been highlighted here, such as
semi-exclusive reactions, studies of the proton fragmentation region, heavy
quark studies, and odderon/pomeron interference.

I thank my collaborators, particularly Johan Rathsman, Carlos Merino, Paul Hoyer, 
Stephane Peigne, and Markus Diehl, for
many helpful discussions.  I also thank the organizers of this meeting at the
Indiana University Cyclotron Facility for stimulating and organizing this
interesting workshop.


\end{document}